# Self-verifiable LLM-driven science and why it needs human researchers

Michael G. Medvedev[1]*, Nikolai V. Krivoshchapov[1]*, Stepan N. Ostarkov[1]*

[1]N.D. Zelinsky Institute of Organic Chemistry of Russian Academy of Sciences, Moscow, Russian Federation

## Abstract

Language models are now used throughout all sciences – for reading literature, extracting numbers from tables and spectra, analyzing data, drafting conclusions – and their answers arrive in the same polished confident tone whether they are right or wrong, so a result has to be re-derived by a human researcher before it can be relied on. The cost of that re-checking, not the ability to generate results, is what currently limits AI-assisted research. In the field of mathematics, this problem is solved by proof assistants like Lean 4 and Coq that let any formal statement and its proof be written down and checked — a perfect match for LLMs, which are now actively used to search for proofs with every step automatically validated, rather than relying on a human to judge the result. In this paper, we propose an extension of this approach to non-mathematical sciences. We describe an AI agent harness that aims to make LLM's scientific reasoning verifiable: the starting assumptions are proposed by an LLM and reviewed by a human, and everything built on top of them is checked by a machine with Lean 4. We illustrate how this mechanism works in practice on several examples and our results show that scientific arguments in these tasks can be expressed as formal proofs from explicit assumptions, including experimental results, known theoretical statements, and additional assumptions, produced by the agent and accepted by the human scientist. We suggest that this approach could change how scientific research is conducted across all sciences.

## Introduction

Large language models (LLMs) have recently entered many scientific workflows, but researchers have no way to separate a supported answer from a misread or invented one except by redoing the work — which is why the limiting factor in AI-assisted science is no longer generation but verification.[1,2]

Before AI can be reliably used to drive scientific research, three major drawbacks must be addressed. First, the correctness of facts, reasoning, and computations must be controlled. In our recent review on AI in chemistry we have discussed[3] the theoretical limitations which make hallucinations unavoidable[4] and their real-life consequences. Second, in scientific research, field-specific rules are commonly implied rather than documented — significance thresholds, admissible transformations, definitional choices, units, sampling assumptions — and a checker that does not carry them cannot see them. For example, under some abundant circumstances 40 plus 40 might equal 100 (in the gym, since there is also a bar). Third, a result that cannot be reliably re-derived from its stated premises is not a scientific result, whereas the output of an LLM is stochastic by construction. To use AI for doing science we must treat these drawbacks not as defects that will be trained away but as properties to be managed.

Great progress in managing and taming LLM deficiencies has been made in software development,[5] where the ultimate goal of a project can usually be formulated as a set of unit and end-to-end tests; these tests then serve as a clear invariant, the job is either done correctly or it is not, which makes it possible to set up agent loops that achieve the goal without human intervention. These acceptance criteria are a part of AI harnesses, that are also responsible for management of LLM context and providing access to the Internet, file system, outputs of terminal commands and other tools so that hallucinations cannot spoil the result.

Remarkably, thanks to the Curry–Howard correspondence and the dependent type theory,[6,7] the process of constructing formal mathematical proofs can also be turned into a kind of programming task with automated machine proof checks being the ultimate "end-to-end" test.[8] In October 2026 OpenAI released hundreds of mathematical results produced by an internal frontier model together with Lean 4 formalizations of many of the proofs, which is the largest demonstration to date that this end-to-end test can be applied to claims no human has checked by hand.[9]

Motivation for the current work is to adopt the convenience of machine verification for other sciences. This task, however, is not that simple: mathematics is the only field where theories are derived from predefined sets of axioms, thus, mathematical theorems can be assigned a canonical meaning and stated within a single framework.[10]

Claims that a treatment effect is not significant at a chosen level, or that an experimental spectrum is consistent with the calculated one, has no canonical mathematical model: units, probability distributions, conventions and sampling assumptions can vary, and the checker must be aware of them, nevertheless. The strictness of formal verification languages pushes all sensitive decisions to the boundary of the formal system (Fig. S1): to the translation of the claims, prior and experimental data from human language to Lean.

In this paper, we propose an approach for automated verifiable scientific research (Fig. 1), where AI researcher generates assumptions absent in the human researcher input, and the human researcher approves or rejects each of them. The reasoning that follows from the approved assumptions is done in Lean 4 by AI researcher, supplemented by additional assumptions if required, and checked by Lean 4 kernel. After which the human researcher again reviews the results and the added assumptions. We illustrate this mechanism on several examples including a Bayesian data analysis, in each of them the researcher is prompted with several questions to clarify the construction of the model and everything else is machine-checked. The choice of assumptions remains a human judgment, but the harness reduces the human role to the planning stage, leaving all subsequent reasoning to be executed automatically.

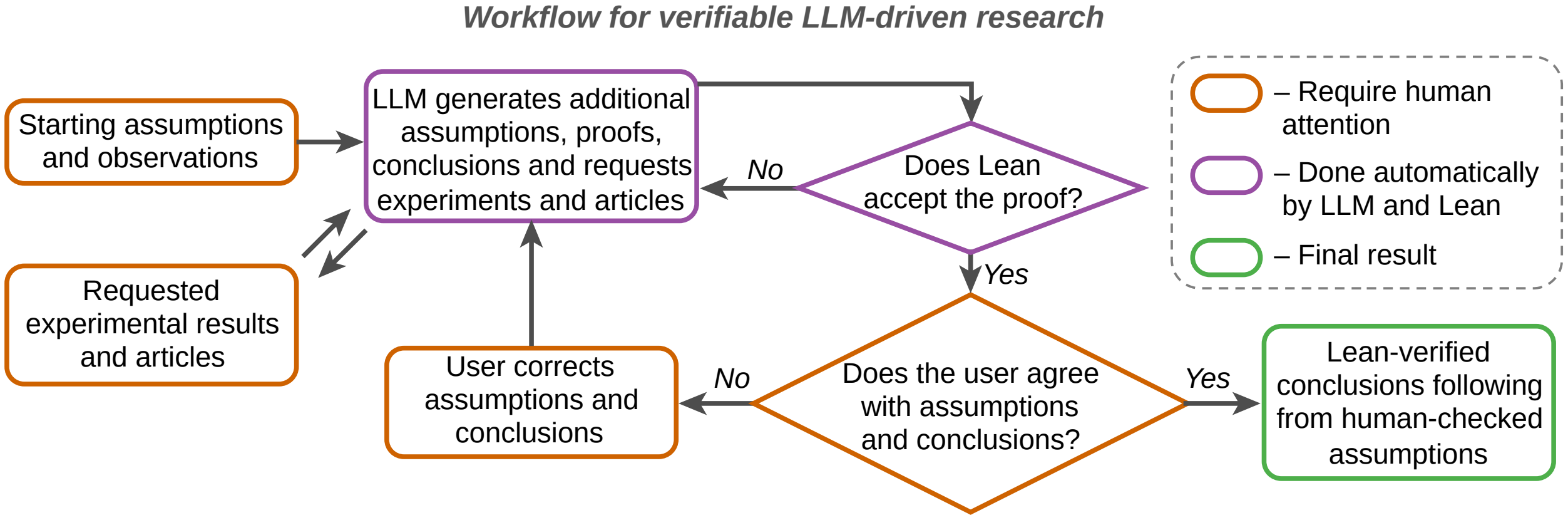

Figure 1. System that aims to eliminate the need for humans to manually verify the entire work generated by AI agent. Instead, human attention (orange) is required only to provide starting assumptions and to review the LLM's proposed assumptions and conclusions (purple).

## Pitfalls of using formal proofs in non-mathematical sciences

To illustrate the difference between an informal reasoning and formal proof, consider a box of 50 balls, from which 10 black balls have been drawn. Given the mean and variance of the proportion of black balls, one can use Bayes' rule to derive the probability of the next ball being black. A short derivation in natural language (as shown on Fig. 2, left) may be convincing to a reader, but it leaves some important assumptions implicit, including the family of the prior distribution and the rule that each drawn ball is kept out of the box. Making these assumptions and the intermediate steps explicit gives the longer formal argument in Fig. 2, center, which can be converted into the corresponding Lean 4 code (see Fig. 2, right).

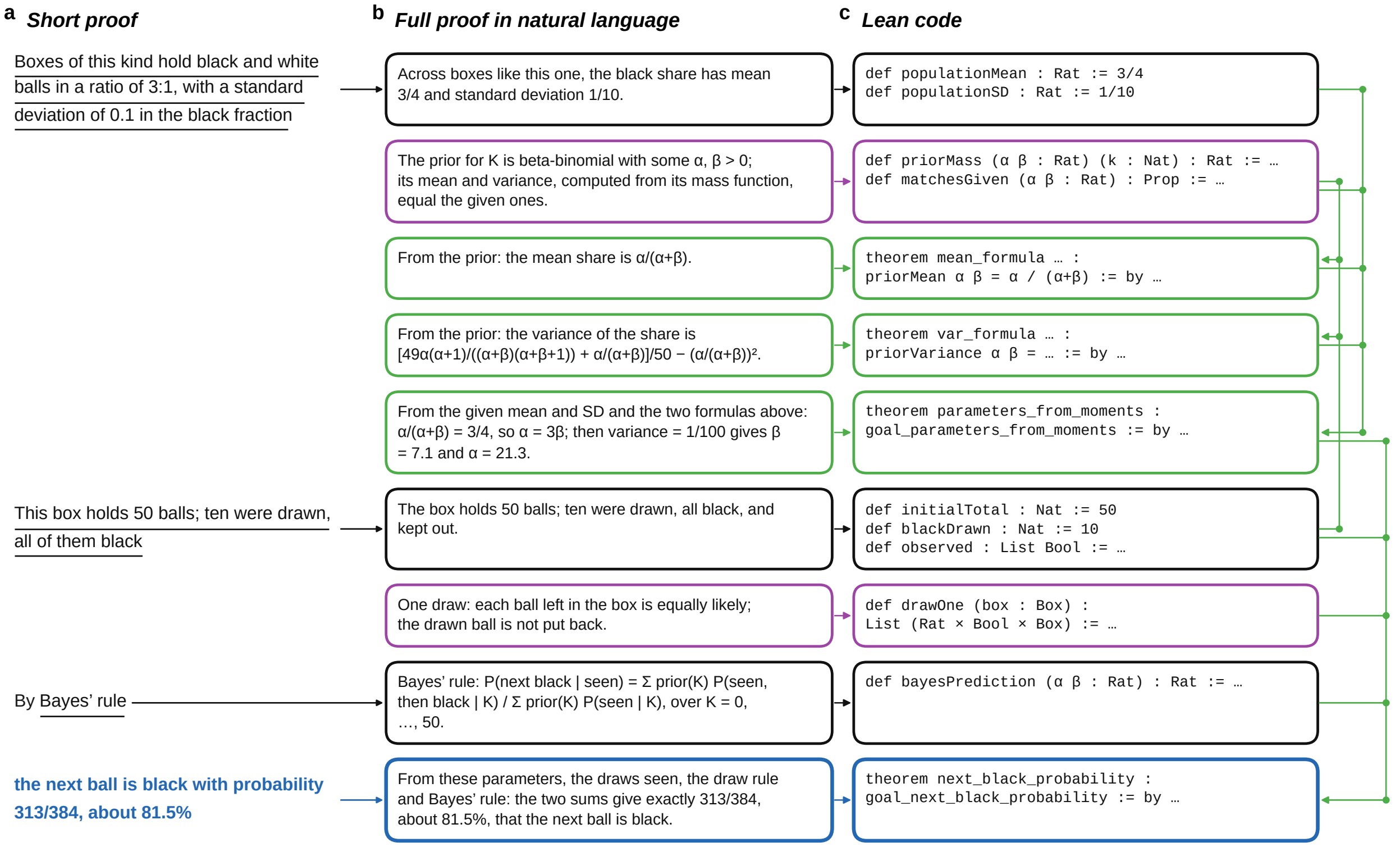

Figure 2. Illustration of the difference between formal and informal proof. Left (a) – an informal reasoning chain that gets an LLM to correct solution. Center (b) – the same proof expanded to be formally correct. Right (c) – the corresponding Lean code. Color shows the role of statements in the proof: black blocks are initial assumptions, purple blocks contain guards (assumptions explicitly added to make the proof formal), green blocks are intermediate lemmas, and blue blocks are conclusions.

In such proofs, we distinguish assumptions (statements about the world, which are taken as true), conclusions (results we aim to derive and use in further reasoning), and lemmas (intermediate propositions that help to reach the conclusion). Lean's proof checker guarantees that an accepted proof establishes its conclusion from the assumptions. Therefore, to assess the AI-derived and Lean-checked results, the scientist should only verify that those assumptions reasonably describe the world around them and that the conclusion is equivalent to the original question.

However, Lean is an engine for deduction: given assumptions and lemmas it calculates whether the conclusion follows from them (Fig. 3, top). However, most of the scientific theories are not direct conclusions from the experiments; on the contrary, the experimental results are predicted from the theory, and the theory that has predicted the experiment the best is considered adequate (Fig. 3, middle). Claims of this kind cannot be formally proven no matter how much data points a scientist has and become formalizable only if we take some additional assumptions (e.g. probability model) to be true (Fig. 3, bottom).

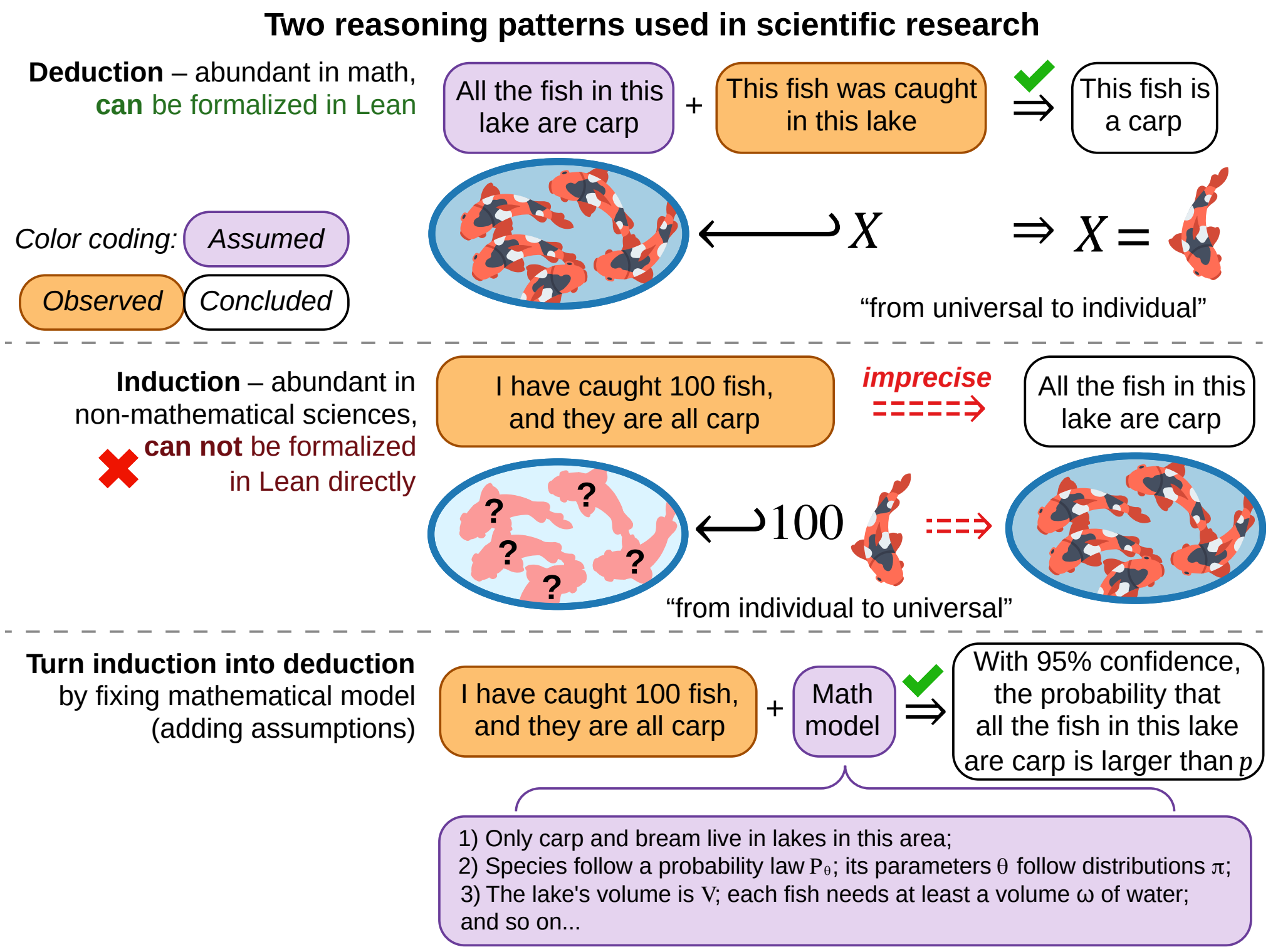

Figure 3. Lean deduces whether a conclusion follows from given statements (top). But scientific research often works in reverse: theory is a generalization of the observations, not a direct conclusion from them. Such claims cannot be formally proven regardless of data amount (middle). Formalization can be achieved if additional statements are assumed, e.g. probability model (bottom).

Verification of all the underlying assumptions (purple in Fig. 3), in particular those proposed by a machine, is crucial for the conclusions to be reliable and is precisely the step that cannot be delegated to a machine – an agent can propose a model and the assumptions it needs, but a human scientist must confirm them. Recent works have approached this boundary from different sides, by reconstructing the missing assumptions explicitly and testing that a proof really needs the premise it claims to use,[11] and by binding the facts a proof rests on to attested, auditable sources.[12] However, to the best of our knowledge, no automatic research system so far has handed the researcher the model and its assumptions for approval and let the machine do everything above.

## Experiment design and results

To show that the proposed human-in-the-loop workflow is feasible and practical, we designed a series of experiments on real scientific reasoning tasks. Each experiment began with a task written down in natural language, passed to the terminal agent (Claude Sonnet 5.5 in Claude Code, or GPT-6.1 Sol in Codex), isolated in a fresh Docker container with the tools giving it an access to its working directory, ability to run shell commands and search the Internet, and request validation of assumptions from a human scientist.

The agent was instructed to formulate a mathematical model, explicitly distinguish supplied information from additional assumptions it made, and derive an answer supported by Lean 4 proof. We also advised agent to derive the model parameters from experimental data and try to include as much of the proof as possible in the intermediate lemmas. It could request further information, consult literature, and revise its model when necessary. Proof attempts were submitted to Lean, which feedback guided subsequent revisions. Once proof passed the formal Lean checks, its assumptions and conclusions were extracted and presented to the human researcher for review as Lean code. If they were rejected, the agent was prompted to review the proof, conclusions and assumptions according to the feedback. Every attempt, checker response and human decision were retained in the experiment record.

We evaluated the system on four tasks: predicting the probability of the next ball drawn from container to be black, estimating fish population in a lake (in this task LLM was allowed to request experiments, which results were derived from a simple simulation and passed back to the model), identifying the best time to pour salt in the heating water, and predicting how a cup of coffee will cool (See Table 1). Two independent AI agents worked on each question, starting from the same written model and data and without sight of one another's work. Both agents obtained correct answers with an accepted proof for all four tasks. Full experimental records are available in SI.

**Table 1. Scientific reasoning tasks and verification.**

| Task | Balls in a box | | Fish in a lake | | Boiling water and salt | | Coffee cooling | |
|---|---|---|---|---|---|---|---|---|
| **Task description (see SI for full version)** | Ten black balls have been taken from a box of 50. Across similar boxes, the black fraction has mean 0.75 and SD 0.1. Predict the next draw. | | Identify fish species and infer their population sizes in an unstudied lake, using requested non-destructive observations. Request experiments if you need them. | | Determine when to add salt while heating one liter of water initially at 20 °C, to reach boiling sooner. | | Determine when coffee initially at 90 °C reaches 60 °C, choosing a cup and thermal model. | |
| **Expected answer** | An updated Bayesian probability for the next black ball. | | Species occurrence and probability distributions for species counts and their total. | | A temperature or time for salt addition that minimizes boiling time. | | A prediction of the time to reach 60 °C from a thermal model. | |
| **Expected assumptions** | A composition prior and a rule for drawing without replacement. | | Models of species occurrence, abundance and detection; links from surveys to the lake population. | | Salt composition and amount; heat input, heat losses and material properties. | | Cup geometry, room conditions, heat capacity and a heat-loss law. | |
| | **Claude Sonnet 5.5** | **GPT-6.1 Sol** | **Claude Sonnet 5.5** | **GPT-6.1 Sol** | **Claude Sonnet 5.5** | **GPT-6.1 Sol** | **Claude Sonnet 5.5** | **GPT-6.1 Sol** |
| **Total lean verifier passes** | 8 | 6 | 43 | 2 | 27 | 14 | 16 | 14 |
| **Total human verifier passes** | 2 | 2 | 6 | 2 | 5 | 5 | 1 | 4 |
| **Full proof, lines** | 363 | 289 | 428 | 207 | 329 | 273 | 115 | 273 |
| **Human-checked part of the proof, lines** | 23 | 24 | 97 | 39 | 41 | 27 | 25 | 16 |

## Conclusions

Here we have described a harness in which an LLM proposes the mathematical model and the starting assumptions to answer the given scientific question, and a human scientist approves or rejects each of them, while the correctness of model's logic is tested in the background by Lean 4. We show that this harness works on all four example tasks for both off-the-shelf agents based on Claude Sonnet 5.5 and GPT-6.1 Sol, which ran independently. Across these tasks human verifier read no more than 97 lines of assumptions and conclusions (typically ~30), while the kernel checked up to 428 lines of proof (typically ~300).

A scientific proof shows that the results follow from the model and the data as they were written down; if those are wrong, the results might be incorrect as well. Thus, human researchers of the future will require solid world models to judge correctness of the machine-proposed assumptions and will need to be fluent in Lean 4 or another formal scientific language, which will be accepted by the scientific community. Databases containing assumptions certified by experts in the respective fields connected[12] to self-verifying scientific harnesses can further reduce the workload required from human researchers.

**Code and data availability**

Our harness, along with the prompts and full traces as machine-readable JSON or human-readable HTML are available in the GitHub repository:
https://github.com/TheorChemGroup/Self-verifiable_CyberScientist.

Full agent traces may be also viewed in the web interface by the following link:
https://theorchemgroup.github.io/Self-verifiable_CyberScientist/traces/

**Author contributions**

M.M.G. conceived the main idea and revised the manuscript. K.N.V. conceptualized the idea and wrote the initial draft. O.S.N. created the harness, conducted the experiments, and wrote the experimental section.

**Declaration of competing interests**

The authors declare no competing interests.

**AI use statement**

While writing the code, article text and interpreting the results, authors used several AI-based instruments, including GPT-6.1 Sol *via* Codex and Claude Opus 5.5 *via* Claude Code. All authors take complete responsibility for all the results and claims in the article. The final version of the article was reviewed and polished by all authors.

# Self-verifiable LLM-driven science and why it needs human researchers

Michael G. Medvedev[1]*, Nikolai V. Krivoshchapov[1]*, Stepan N. Ostarkov[1]*

[1]N.D. Zelinsky Institute of Organic Chemistry of Russian Academy of Sciences, Moscow, Russian Federation

## Supplementary information

Our harness, along with the prompts and full traces as machine-readable JSON or human-readable HTML is available in the following GitHub repository: https://github.com/TheorChemGroup/Self-verifiable_CyberScientist.

We used four scientific reasoning scenarios: Bayesian prediction for balls drawn from a box, inference of fish species and population sizes in a lake, the timing of salt addition when heating water, and coffee cooling (see Table S1). Each question was supplied in natural language to a terminal agent. The agent proposed a mathematical model, identified additional assumptions, and attempted to derive an answer in Lean. A human researcher reviewed the assumptions, definitions and conclusions of machine-checked submissions and could require further revisions.

The recorded clients were Codex CLI 0.166.0 and Claude Code 2.1.325. The requested models were gpt-6.1-sol and claude-sonnet-5-5, respectively, with medium reasoning effort.

Formal checking used Lean 4.35.0 and a pinned Mathlib dependency set.

Each invocation started a fresh Docker container with a temporary home directory and client configuration. The agent could write to its own run workspace and temporary directories. Selected harness files, the current task and instructions, the native executable and the Lean installation and library builds were mounted read-only. Other runs, previous solutions, the host's home directory, global CLAUDE.md/AGENTS.md and client configuration, repository history and the Docker socket were not available for the agent. Internet access was available for the native client and literature research.

The paired salt and coffee runs used limits of 4 GiB memory and two CPUs per scientist/checker job. Lake-runs initially used 3 GiB and two CPUs. Claude's sixth lake cycle ended with exit code 137 during a Lean check; Docker recorded two out-of-memory events for that container. The partial work and trace were retained, and the seventh cycle resumed with 6 GiB and the same two CPUs.

**Table S1. Full task formulations and reviewed models.**

| Task | Balls in a box | Fish in a lake | Boiling water and salt | Coffee cooling |
|---|---|---|---|---|
| **Task description** | I have a box that initially contained 50 balls, each either black or white. I drew 10 balls, all of which were black, and did not return them to the box. There are now 40 balls remaining. Other boxes of the same kind also initially contain 50 black and white balls. Across these boxes, the fraction of black balls has a mean of 0.75 and a standard deviation of 0.1. I do not know the original number of black balls in my particular box. Using Bayesian statistics, what is the probability that the next ball drawn from the remaining balls will be black? | We are standing near an unnamed lake in North Karelia, eastern Finland. Your goal is to determine which fish species are present, propose a probability distribution for the number of individuals of each species, and derive the probability distribution of the total fish population. No surveys, measurements or catches have been performed. You may propose observations or experiments and request their results. Fish must not be caught or removed from the lake. | I want to boil saline water. I have poured 1 litre of water at 20 degrees Celsius into a pot and started heating it on a stove. I will add salt, and want the final salted water to reach its boiling point as quickly as possible. The salt quantity, pot, heater and room conditions have not otherwise been specified. At what water temperature, or after how much heating time, should I add the salt to make the salted water boil sooner? | I have a cup of coffee initially at 90 degrees Celsius and want to know when it will cool to 60 degrees Celsius. Choose a reasonable cup shape and size yourself, explicitly identifying these as chosen scenario details rather than measured facts. You may request measurements of temperature, cup geometry or surrounding conditions. No time-temperature measurements or cooling coefficients have yet been supplied. How long will the coffee take to cool to 60 degrees Celsius? |
| **Expected answer** | An updated Bayesian probability for the next black ball. | Species occurrence and probability distributions for species counts and their total. | A temperature or time for salt addition that minimizes boiling time. | A prediction of the time to reach 60 °C from a thermal model. |
| **Expected assumptions** | A composition prior and a rule for drawing without replacement. | Models of species occurrence, abundance and detection; links from surveys to the lake population. | Salt composition and amount; heat input, heat losses and material properties. | Cup geometry, room conditions, heat capacity and a heat-loss law. |

| Task | Balls in a box | | Fish in a lake | | Boiling water and salt | | Coffee cooling | |
|---|---|---|---|---|---|---|---|---|
| | **Claude Sonnet 5.5** | **GPT-6.1 Sol** | **Claude Sonnet 5.5** | **GPT-6.1 Sol** | **Claude Sonnet 5.5** | **GPT-6.1 Sol** | **Claude Sonnet 5.5** | **GPT-6.1 Sol** |
| **Agent's answer (summary)** | 313/384. The beta-binomial parameters a = 21.3 and b = 7.1 and the posterior prediction are derived in Lean. | 313/384. Finite count-prior moments identify a = 21.3 and b = 7.1; the posterior average accounts for the ten removed black balls. | Normalized species-count laws and their total convolution. Posterior-mean bounds are 5870–5880 in the central scenario, 2830–2840 at higher detectability and 17000–17100 at lower detectability. | Normalized species-count and total-count distributions. A floating evaluation gives mean approximately 3388 and interval [2554, 4399]; those decimal evaluations are not certified Lean bounds. | Immediate addition is no slower. Temperature-dependent material calculations establish cooling on dissolution and its timing consequence under the heat-loss model. | Immediate addition is no slower. A fitted capacity family and enthalpy comparison establish cooling on mixing and the timing consequence. | Exact crossing ln(7/4)/k; a formal 1100–2100 s enclosure. A floating evaluation is about 22 min. | Exact crossing (C/G) ln(7/4); a nominal 30–32 min bound and 18–59 min over the stated parameter ranges. |

| Task | Balls in a box | | Fish in a lake | | Boiling water and salt | | Coffee cooling | |
|---|---|---|---|---|---|---|---|---|
| | **Claude Sonnet 5.5** | **GPT-6.1 Sol** | **Claude Sonnet 5.5** | **GPT-6.1 Sol** | **Claude Sonnet 5.5** | **GPT-6.1 Sol** | **Claude Sonnet 5.5** | **GPT-6.1 Sol** |
| **Assumptions used (translated to natural language)** | Beta-binomial prior for the initial black count; supplied count, moments and ten-black observation; uniform sampling without replacement. The general variance and prediction formulas are proved. | Beta-binomial initial count prior; supplied population moments; hypergeometric likelihood for ten black draws; remaining fraction (K−10)/40. Parameters and normalization are derived. | Eight taxa present and exhaustive; complete mapped area; geometric cell-count priors, a discrete density hyperprior and repeated-label likelihood. Calibration-to-field transfer and error approximations remain scenario assumptions. | Eight taxa present and exhaustive; complete mapped area; positive counts supported on 1–10000; declared priors, Poisson repeat encounters/false labels and finite calibration uncertainty. Exact transfer and homogeneous response are assumed. | Salt/water mass ratio 0.02–0.05; Laliberte heat-capacity correlation, dissolution reference data, crystal-capacity and phase bounds. Constant heater input, Newtonian heat loss and conservative mixing. | 10 g NaCl per litre; rational temperature families fitted to Parker data and affine composition dependence; extrapolation to the declared temperature range. Constant heater input, Newtonian heat loss and conservative mixing. | Chosen 250 mL cylindrical cup, radius 4 cm; effective heat-transfer coefficient 25 W/(m²·K); 20 °C ambient; no mug heat storage. The heat-balance equation is assumed and its solution derived. | Chosen cylindrical mug geometry and material bounds; coffee and mug heat storage; exposed areas and capacity derived from components; effective side/top transfer and 20 °C ambient. The cooling solution is derived. |
| **Lean verifier passes** | 8 | 6 | 43 | 2 | 27 | 14 | 16 | 14 |
| **Human verifier passes** | 2 | 2 | 6 | 2 | 5 | 5 | 1 | 4 |
| **Total proof, lines** | 363 | 289 | 428 | 207 | 329 | 273 | 115 | 273 |
| **Human-checked part of the proof, lines** | 23 | 24 | 97 | 39 | 41 | 27 | 25 | 16 |

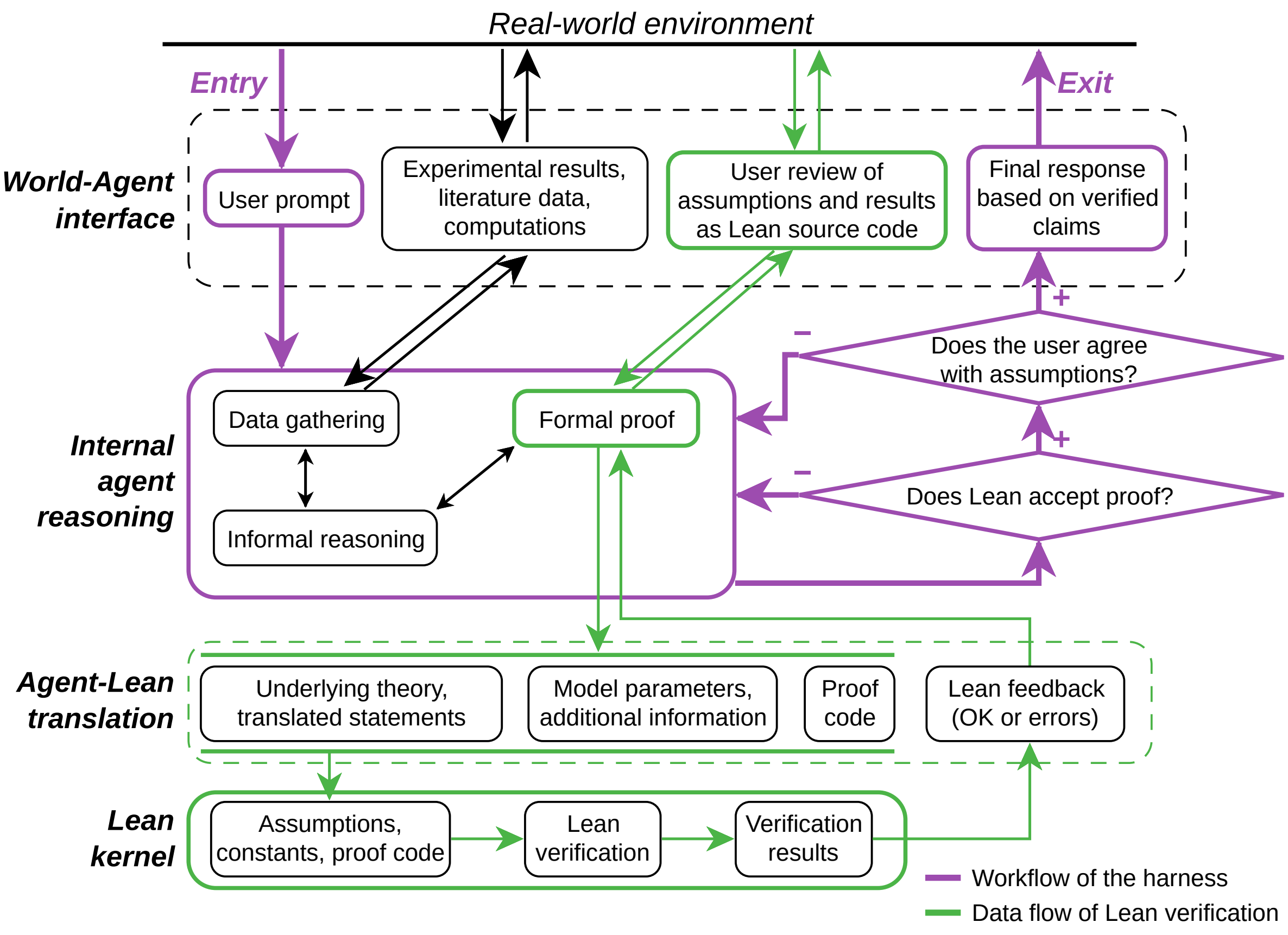

Figure S1. The concept of AI-driven verifiable scientific research. Correctness of verification relies on accuracy of translation from informal text description to formal Lean code (see Agent-Lean translation block).